\documentclass{PoS}

\title{Electroweakino and slepton pair production at the LHC in NLO+NLL with resummation-improved PDFs}

\ShortTitle{Electroweakino/slepton pair production in NLO+NLL with resummation-improved PDFs}

\author{Juri Fiaschi, \speaker{Michael Klasen}\thanks{Work supported by the BMBF under contract
  05H18PMCC1 and the DFG through the Research Training Group 2149 ``Strong and weak
  interactions -- from hadrons to dark matter''.}\\
  Institut f\"ur Theoretische Physik,
  Westf\"alische Wilhelms-Universit\"at M\"unster,
  Wilhelm-Klemm-Stra\ss{}e 9,
  48149 M\"unster, Germany\\
  E-mail: \email{fiaschi@uni-muenster.de}, \email{michael.klasen@uni-muenster.de}}

\abstract{Using parton density functions (PDFs) with threshold-resummation improvement, we consistently calculate higgsino/gaugino and slepton pair production at next-to-leading order and next-to-leading logarithmic accuracy at the LHC. The smaller PDF uncertainty of the global PDF sets is exploited with a factorisation method, which also avoids complications arising in the refitting of threshold-resummation improved PDF replicas in Mellin space. We explicitly take into account the reduction of the scale uncertainty due to the resummation. We show that the resummation contributions in the PDF fits partially compensate the cross section enhancements induced by those in the partonic matrix elements.}

\FullConference{
European Physical Society Conference on High Energy Physics - EPS-HEP2019 -\\
			10-17 July, 2019\\
			Ghent, Belgium}

\begin{document}

\section{Motivation}

Supersymmetry (SUSY) is an extension of the Standard Model (SM) that
is not only well motivated at high scales as the largest possible
extension of the Poincar\'e symmetry, that includes supergravity
and is an essential part of string theory, but also at the weak scale,
where it proposes a solution to the hierarchy, unification and dark
matter problems \cite{Klasen:2015uma}. Recently, simplified models
have been employed to not only focus on the relevant SUSY particle
sectors, but also as prototypes for many more minimal Beyond-the-SM
theories \cite{Fuks:2017rio}. Therefore, the LHC experiments ATLAS
and CMS actively search for SUSY particles in simplified scenarios
and will continue to do so with higher luminosity in the upcoming runs.

To fully exploit the experimental discovery potential, precise
theoretical calculations of SUSY particle production are required.
They have been performed at fixed, next-to-leading order (NLO) and
with resummation methods to all orders at next-to-leading-logarithmic
(NLL) accuracy and beyond for sleptons \cite{Beenakker:1999xh,Bozzi:2006fw},
gauginos/higgsinos \cite{Beenakker:1999xh,Debove:2009ia}, squarks and gluinos
\cite{Beenakker:1996ch,Beenakker:2016lwe} and associated gluino-gau\-gino/higgsino production
\cite{Spira:2002rd,Fuks:2016vdc}. Note that the sensitivity for this last process
at the High-Luminosity LHC has recently been studied \cite{Carpenter:2018ofo}.
The calculations for the electroweak and semi-strong channels have been
made public with the code RESUMMINO \cite{Fuks:2013vua}. They generally
increase the total cross sections and reduce the scale dependence.
In our previous work, the parton density function (PDF) uncertainty remained
at NLO, since it was not reduced by the resummation in the hard matrix elements
of the SUSY particle production processes.

\section{Formalism}


PDFs with resummation improvement are now available with the NNPDF3.0
set \cite{Bonvini:2015ira}. However, as NLO+NLL calculations could not
be used for all processes employed in typical global PDF fits, the
resummation-improved fit had to be performed with the reduced data set
of deep-inelastic scattering, Drell-Yan and top-pair production data.
To consistently show the impact of the resummation contributions, an
NLO analysis of the same data set has also been performed.

The NNPDF analyses employ a replica method that is known to be problematic
in Mellin space, where PDFs are required at all values of $x$. In particular,
outliers can produce unphysical (e.g.\ negative) cross sections. Also, due
to the reduced data set the resummation-improved PDF analysis presented a
larger, not smaller uncertainty than the global analysis.


To avoid both shortcomings, a method based on ratios of NLL and NLO cross
sections with reduced and global PDF fits can be employed \cite{Beenakker:2015rna}. 
By defining a $K$-factor as
\begin{equation}
 K =
 \frac{\sigma({\rm NLO+NLL})_{\rm NLO~global}}{\sigma({\rm NLO})_{\rm NLO~global}}
 \cdot
 \frac{\sigma({\rm NLO+NLL})_{\rm NLO+NLL~reduced}}{\sigma({\rm NLO+NLL})_{\rm NLO~reduced}},\nonumber
\end{equation}
(approximate) NLO+NLL cross sections with NLO+NLL PDFs can be obtained with
\begin{equation}
 \sigma({\rm NLO+NLL})_{\rm NLL+NLO~global} = K \cdot \sigma({\rm NLO})_{\rm NLO~global}.\nonumber
\end{equation}
This allows also to obtain (approximate) NLO+NLL global PDF errors by
varying the NLO global PDFs in $\sigma({\rm NLO})_{\rm NLO~global}$, and at the
same time to eliminate replicas with unphysical behaviour. The renormalization
and factorization scales are still varied with the seven-point method directly
in $\sigma({\rm NLO+NLL})$. The total theoretical uncertainty is obtained by
adding PDF and scale errors in quadrature. This method can not only be applied
to total cross sections $\sigma$, but also to distributions in the invariant
mass $M$ of the produced SUSY particle pair $d\sigma/dM_{\tilde{\ell}\tilde{\ell},
\tilde{\chi}\tilde{\chi}}$.


We apply this method to SUSY particle production in simplified scenarios of
the Minimal Supersymmetric Standard Model (MSSM). For sleptons, we assume
pure left-handed mass-degenerate selectrons and smuons and right-handed or
maximally mixed staus. For neutralinos and charginos, we assume a natural
SUSY spectrum with Higgsino masses below 1 TeV and a compressed spectrum
of $m_{\tilde{\chi}^\pm_1}\simeq m_{\tilde{\chi}^0_2}\simeq m_{\tilde{\chi}^0_1}$
as well as heavier gauginos.
The processes that are implemented in RESUMMINO include the electroweak
processes $pp\to \tilde{\ell} \tilde{\ell}^*$ ($\ell=e,\mu$), $\tilde{\tau}_R
\tilde{\tau}_R^*$, $\tilde{\tau}_1\tilde{\tau}_1^*$; $pp\to \tilde{\chi}^\pm_i
\tilde{\chi}^\mp_j$, $\tilde{\chi}^\pm_i\tilde{\chi}^0_j$, $\tilde{\chi}^0_i
\tilde{\chi}^0_j$ ($i,j=1,2$); the semi-strong processes $pp\to\tilde{g}
\tilde{\chi}^{\pm,0}_j$, $\tilde{q}_i\tilde{\chi}^{\pm,0}_j$ (in progress);
and the GUT processes $pp\to Z'\to \ell\bar{\ell}$, $pp\to W'\to \ell\bar{\nu}$.
Since the produced sleptons, charginos and neutralinos decay as
$\tilde{\ell}\to\ell\tilde{\chi}^0_1$, $\tilde{\chi}^\pm_1\to W^\pm\tilde{\chi}^0_1
\to\ell\bar{\nu}\tilde{\chi}^0_1$ and $\tilde{\chi}^0_2\to Z\tilde{\chi}^0_1\to
\ell\bar{\ell}\tilde{\chi}^0_1$, the experimental signatures involve
(sometimes soft) leptons and (possibly moderate) missing transverse energy $\not{\!E}_T$.
We do not include the corresponding branching ratios and efficiencies, since they
are usually taken into account experimentally \cite{Fiaschi:2018xdm}.

\section{Slepton/stau production}

\begin{figure}
 \includegraphics[width=0.49\textwidth]{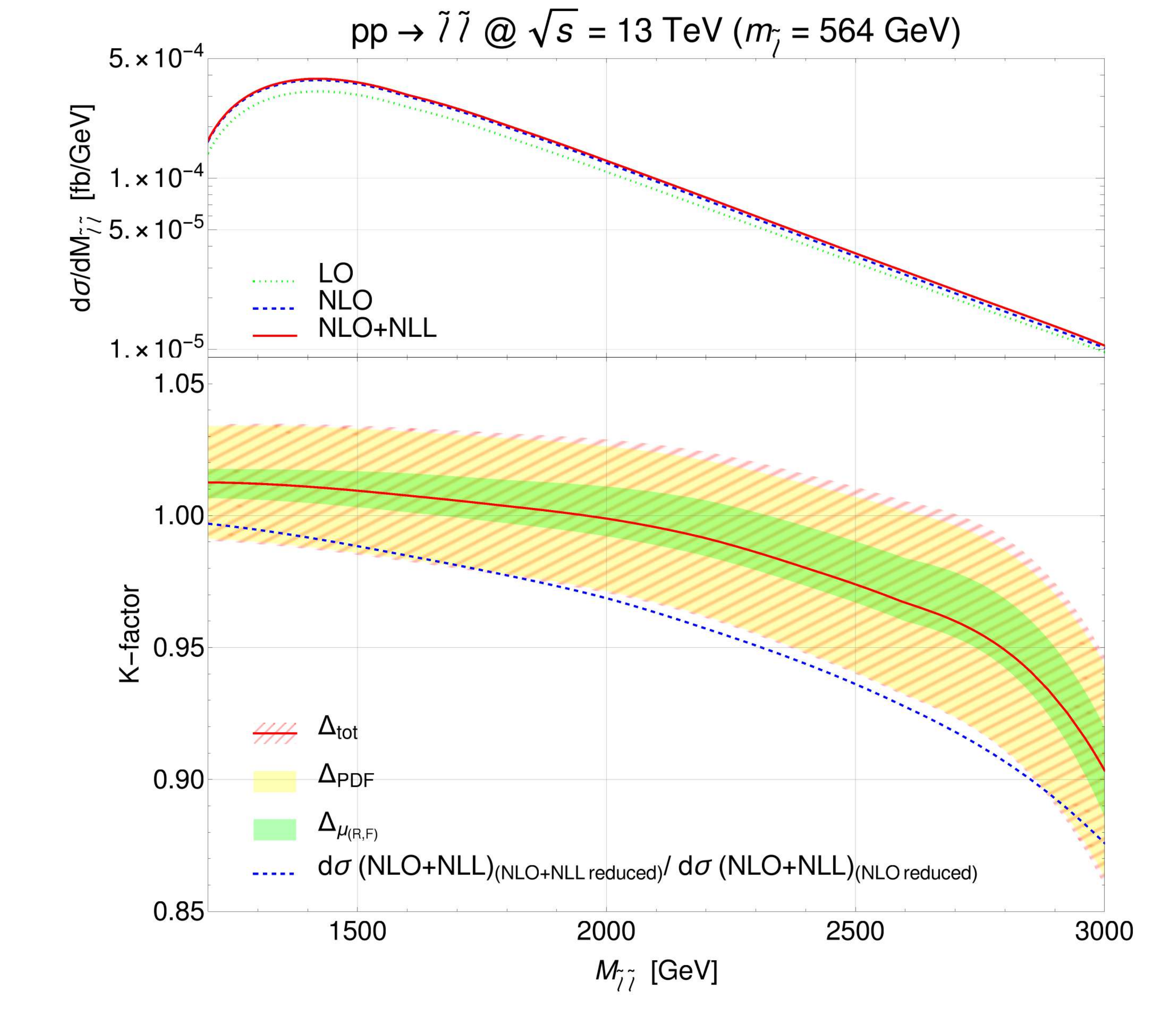}
 \includegraphics[width=0.49\textwidth]{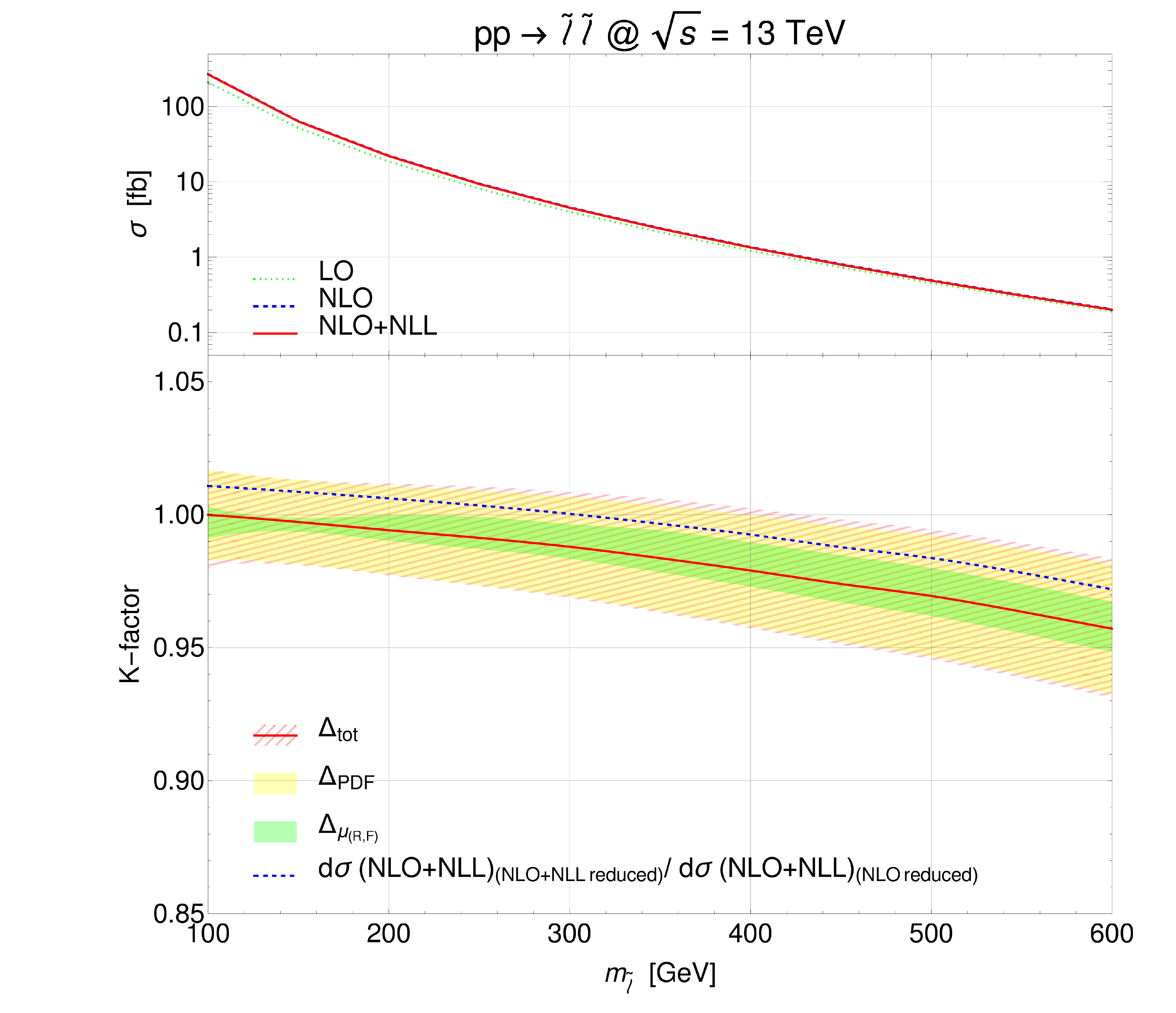}
 \caption{Left: Invariant-mass distributions of left-handed slepton pairs produced
   at the LHC with 13 TeV center-of-mass energy. Right: Corresponding total cross
   sections as a function of the slepton mass. Shown are results at LO (green dotted),
   NLO (blue dashed) and NLO+NLL (red full) together with the corresponding $K$-factors
   and their uncertainties (lower panels). \label{fig:1}}
\end{figure}

For pairs of left-handed sleptons with 564 GeV mass, we show the invariant distribution
in Fig.\ \ref{fig:1} (left) in leading order (LO) (green dotted), NLO (blue dashed) and
NLO+NLL (red full) together with the corresponding $K$-factors and their uncertainties
(lower panel). At the peak, the NLO corrections increase the LO prediction by 16\%,
while the NLL contributions increase it in addition by 2-3\%. In the lower panel, the
red line shows the full $K$-factor, while the blue dashed line indicates the effect of
resummation in the PDFs only. As one can see, they partially compensate the resummation
effects in the matrix elements.

Fig.\ \ref{fig:1} (right) shows the corresponding total cross section as a function
of the slepton mass. In this case, the NLO+NLL cross sections with NLO+NLL PDFs are
smaller by $\sim$ 4\%. Assuming three generations of mass-degenerate sleptons, ATLAS
can now exclude masses up to 700 GeV, while for left-handed selectrons and smuons
individually the limits reach 550 GeV. The sensitivity drops sharply with the mass
difference of sleptons and neutralinos \cite{Aad:2019vnb}.

\begin{figure}
 \includegraphics[width=0.49\textwidth]{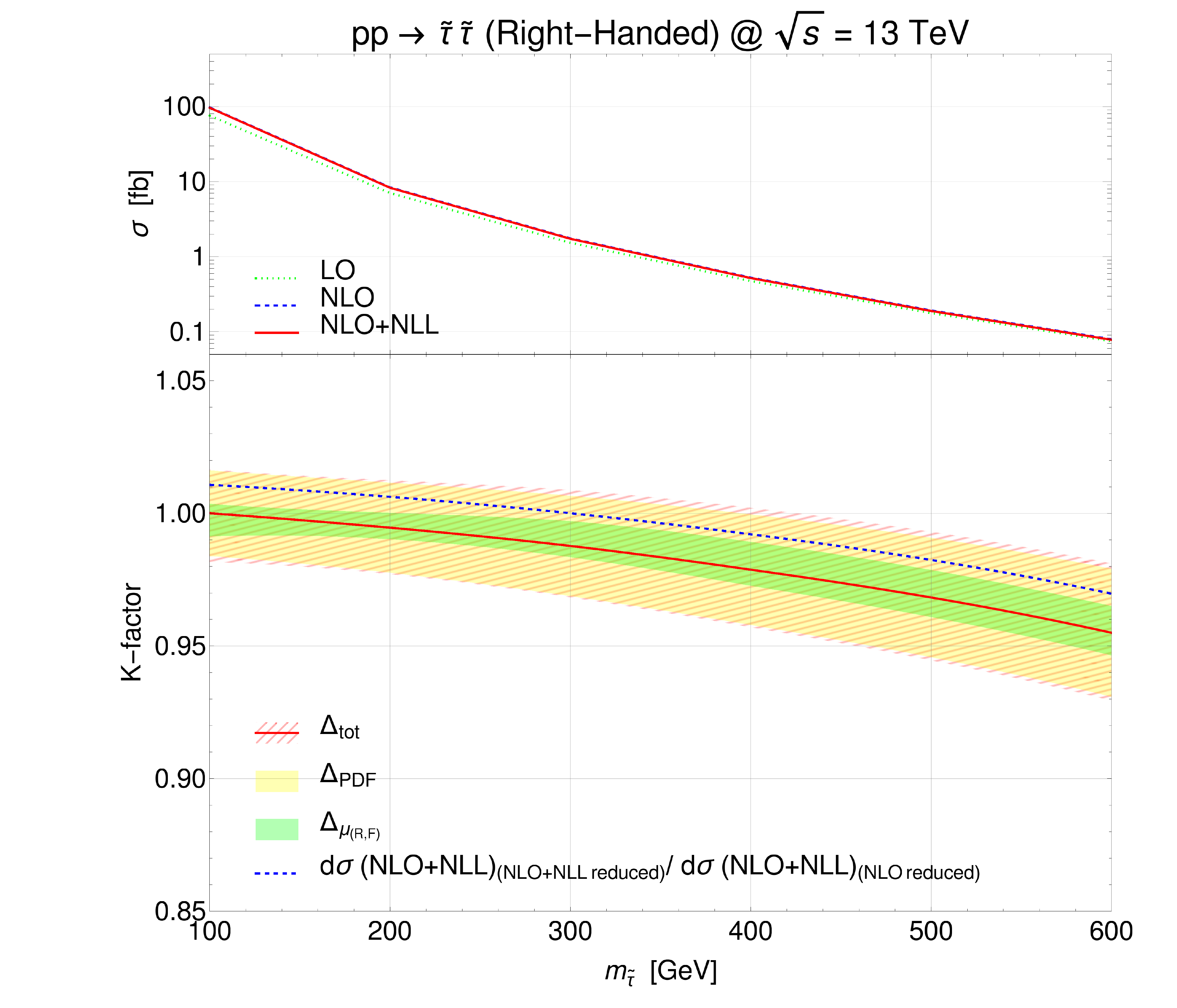}
 \includegraphics[width=0.49\textwidth]{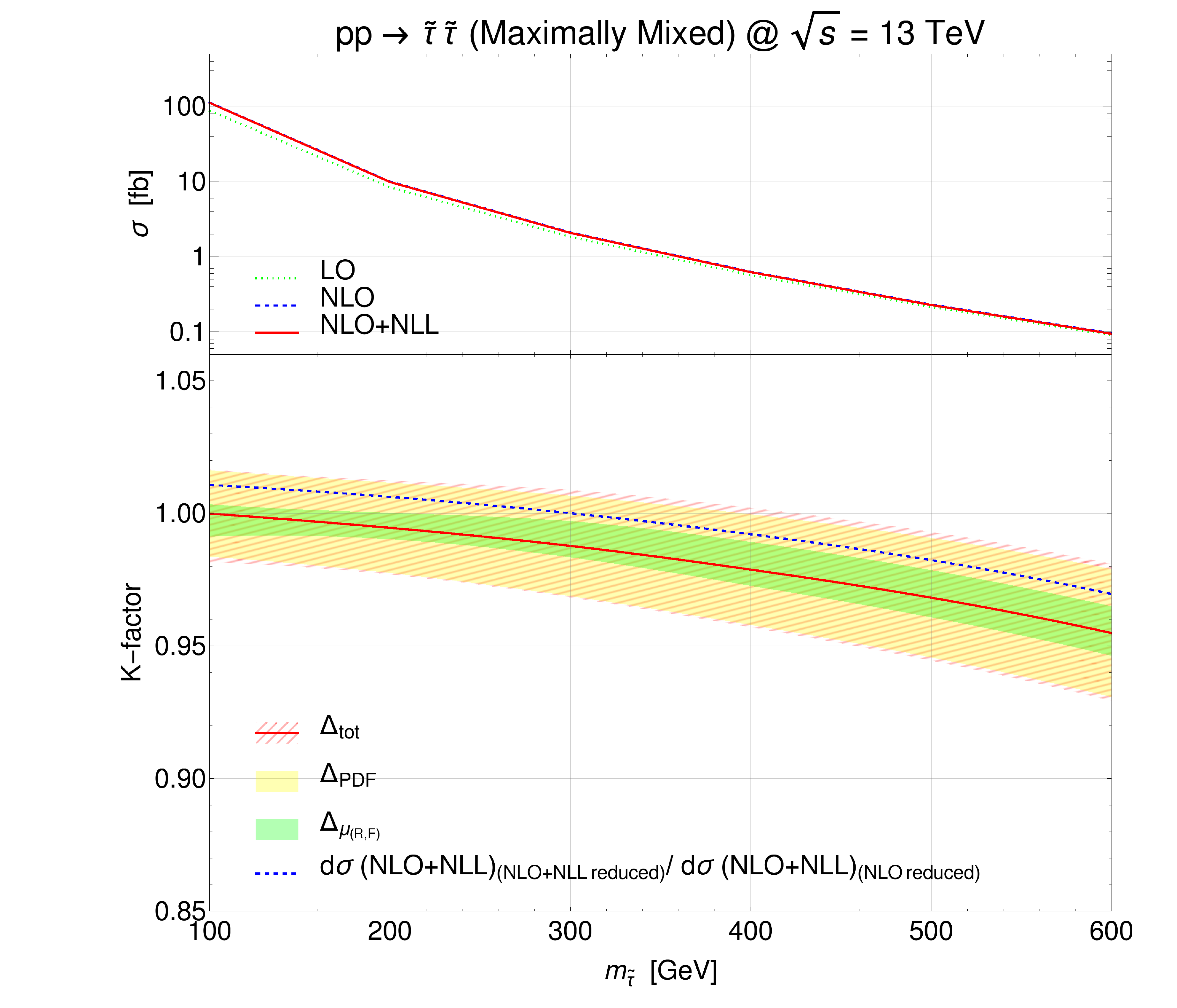}
 \caption{Left: Total cross sections of right-handed stau pairs produced
   at the LHC with 13 TeV center-of-mass energy as a function of the stau
   mass. Right: Same for maximally mixed stau pairs. Shown are results at
   LO (green dotted), NLO (blue dashed) and NLO+NLL (red full) together
   with the corresponding $K$-factors and their uncertainties (lower panels). \label{fig:2}}
\end{figure}

The corresponding total cross sections for right-handed and maximally mixed
stau pairs are shown in Fig.\ \ref{fig:2}.
The total cross sections for right-handed staus (left) are smaller than
those for left-handed staus by almost a factor of three, those for maximally
mixed staus (right) by about a factor of two. The individual corrections
and uncertainties from the higher-order predictions show a very similar
behavior as in the left-handed case.  Current CMS limits on stau masses,
obtained with an integrated luminosity of ${\cal L}=77.2$ fb$^{-1}$, lie
at $m_{\tilde{\tau}}>90$ ... $200$ GeV \cite{CMS:2019eln}.

\section{Higgsino/gaugino production}

\begin{figure}
  \centering
  \includegraphics[width=0.49\textwidth]{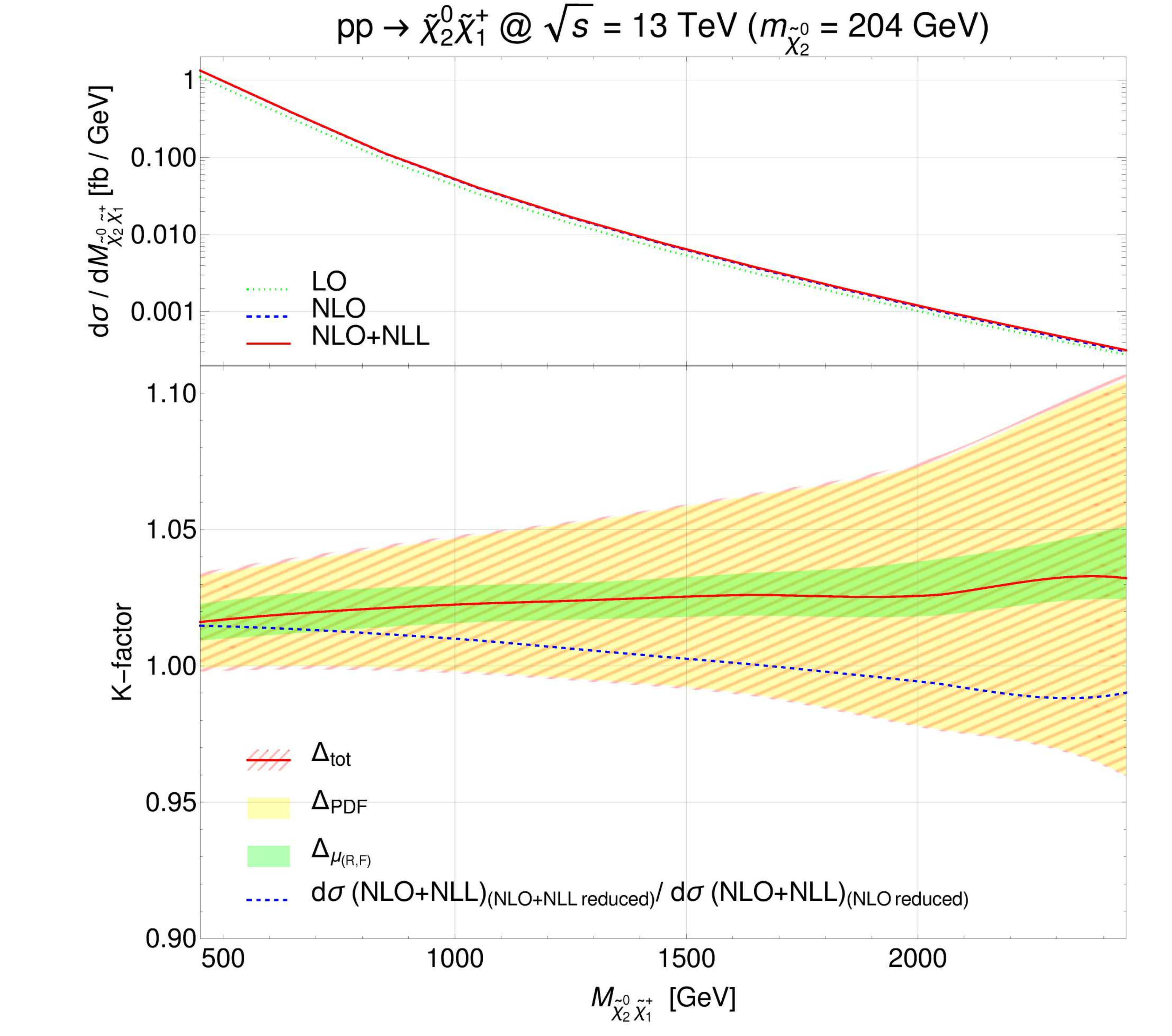}
 \caption{Invariant-mass distributions of light higgsino pairs produced
   at the LHC with 13 TeV center-of-mass energy. Shown are results at LO (green dotted),
   NLO (blue dashed) and NLO+NLL (red full) together with the corresponding $K$-factors
   and their uncertainties (lower panels). \label{fig:3}}
\end{figure}

Turning to electroweakino production, we first show in Fig.\ \ref{fig:3} the
invariant mass distribution for light higgsino pairs of 204 GeV mass.
They are obtained by setting the higgsino mass parameter $\mu$ to 200 GeV
while keeping the bino and wino mass paramters $M_{1,2}$ at 1 TeV. In this
compressed scenario, the mass differences of second-lightest neutralino
and lightest chargino amount to only $m_{\tilde{\chi}^0_2}-m_{\tilde{\chi}^\pm_1}
\approx m_{\tilde{\chi}^\pm_1}-m_{\tilde{\chi}^0_1} \approx$ 5 GeV.
At low invariant mass $M_{\tilde{\chi}\tilde{\chi}}$, the NLO corrections
increase the LO prediction by 20\%, while the NLL contributions increase
it in addition by 1-4\%.

\begin{figure}
 \includegraphics[width=0.49\textwidth]{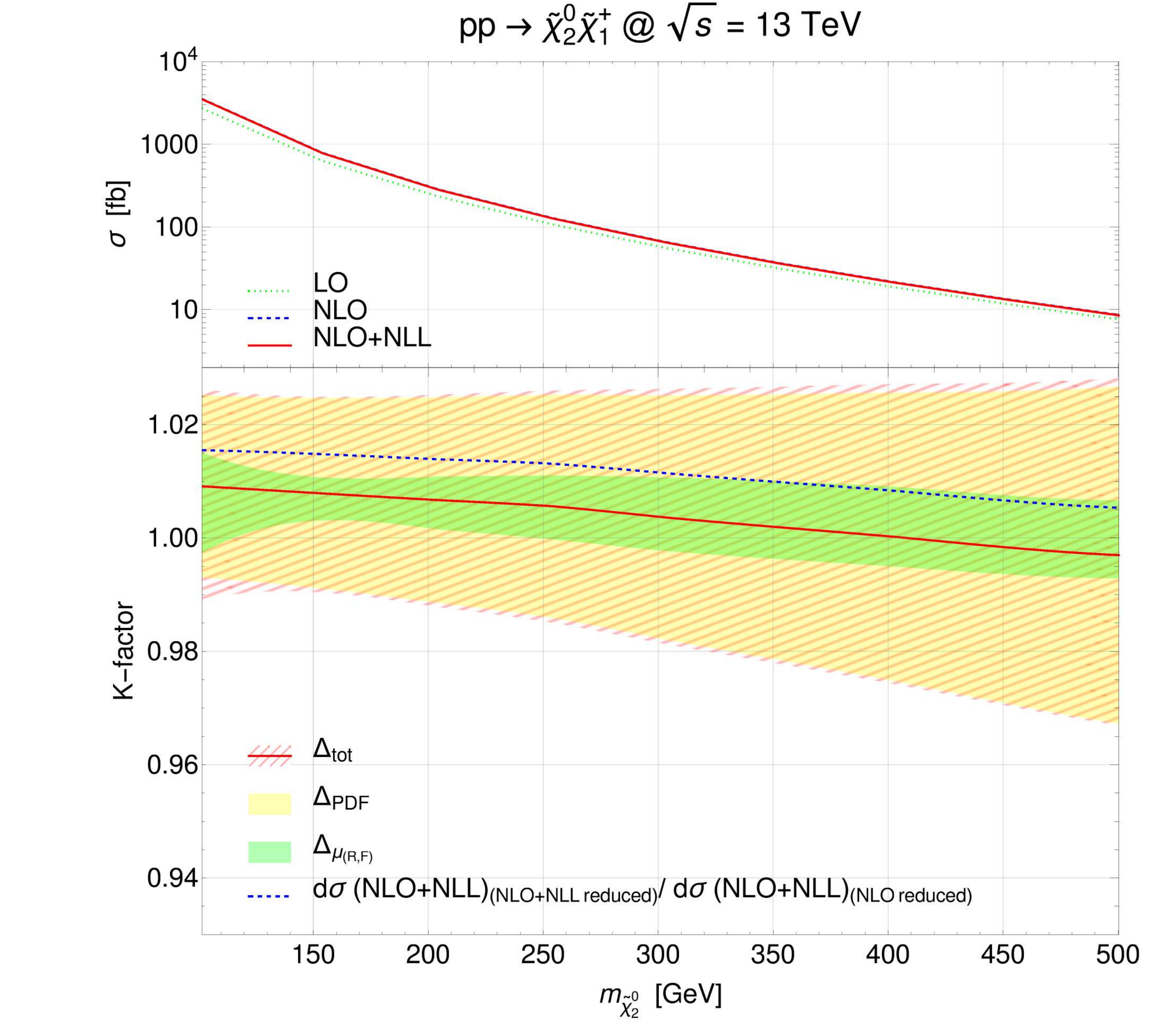}
 \includegraphics[width=0.49\textwidth]{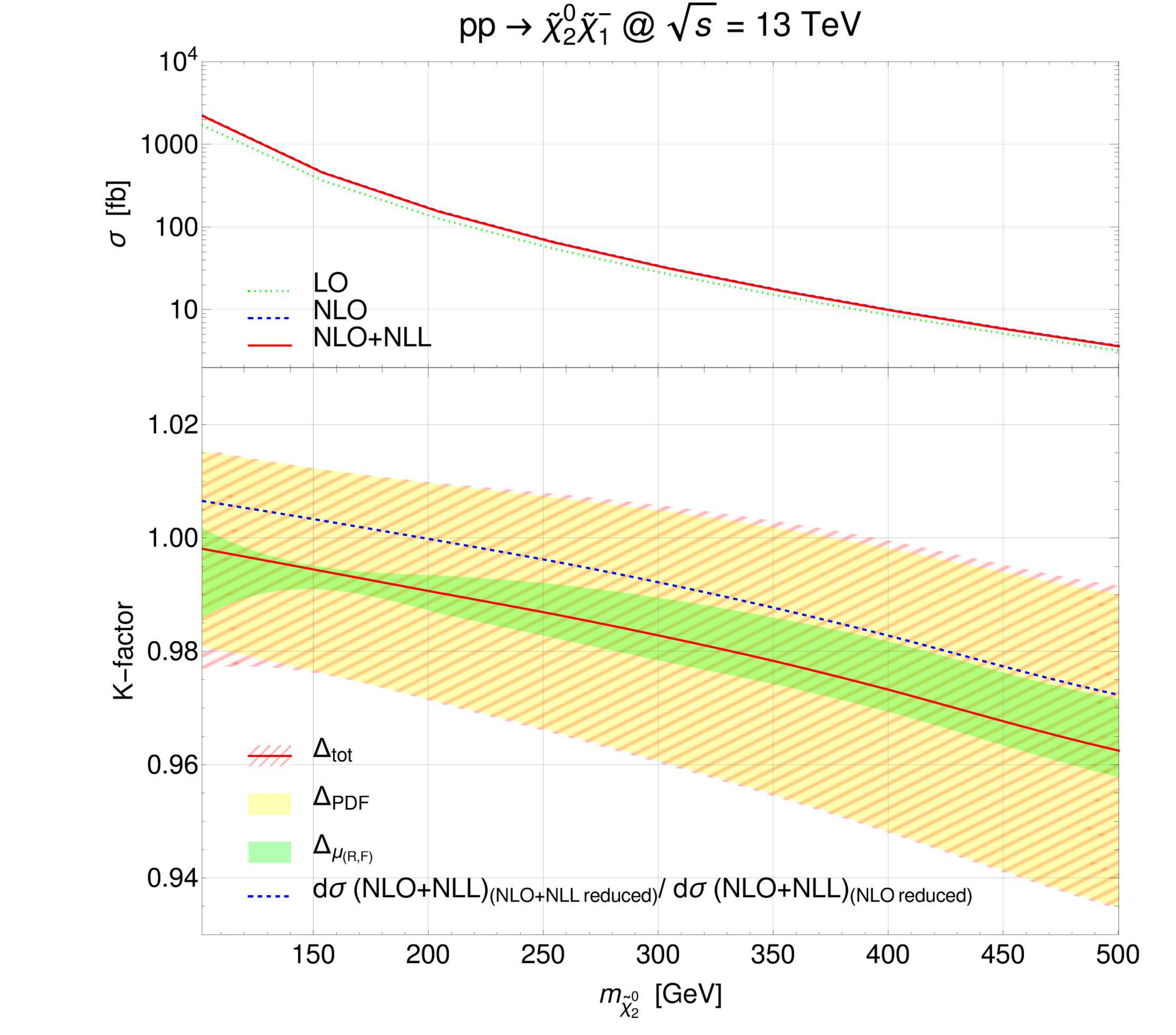}
 \caption{Left: Total cross sections of positive charginos and neutralinos produced
   at the LHC with 13 TeV center-of-mass energy as a function of the higgsino
   mass. Right: Same for negative charginos. Shown are results at
   LO (green dotted), NLO (blue dashed) and NLO+NLL (red full) together
   with the corresponding $K$-factors and their uncertainties (lower panels). \label{fig:4}}
\end{figure}

The corresponding total cross sections are shown in Fig.\ \ref{fig:4}
as a function of the higgsino mass for positive charginos (left) and
negative charginos (right). At a $pp$ collider like the LHC, the former
have of course the larger cross section, which is increased by using the
NLO+NLL PDFs. Negative charginos not only have a smaller total cross
section, but it is also reduced by using NLO+NLL PDFs.

\begin{figure}
 \includegraphics[width=0.49\textwidth]{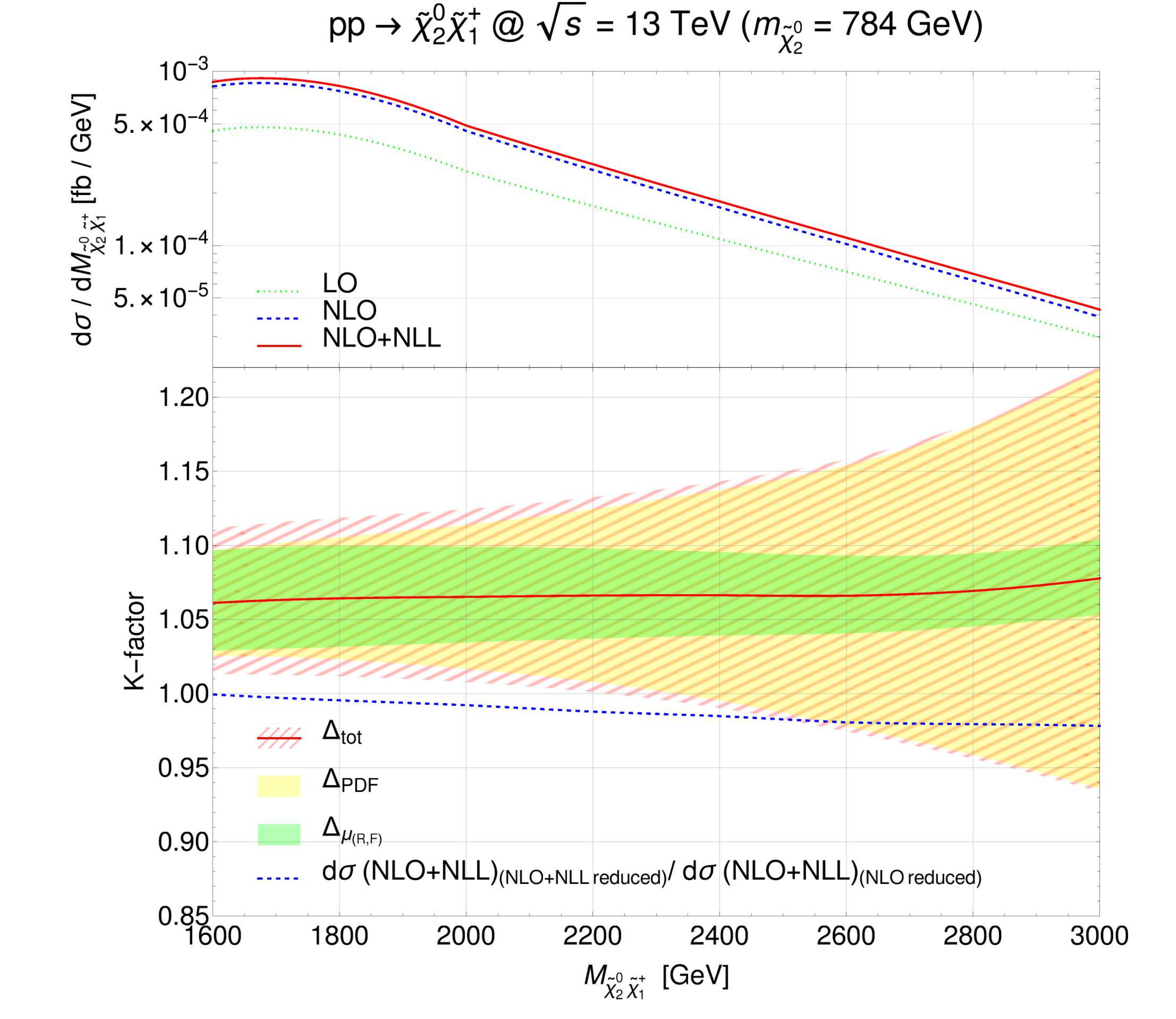}
 \includegraphics[width=0.49\textwidth]{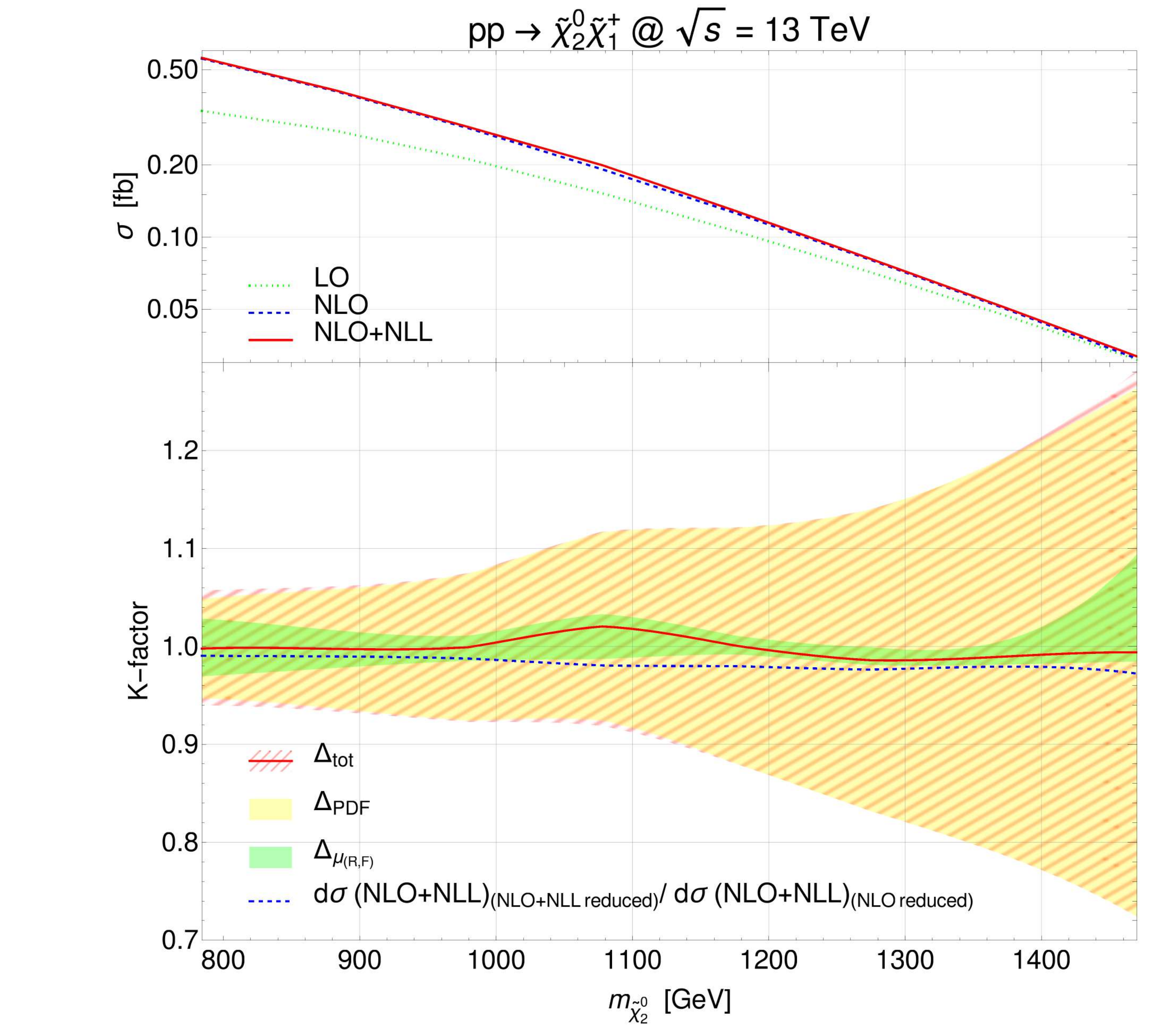}
 \caption{Left: Invariant-mass distributions of heavier gaugino pairs produced
   at the LHC with 13 TeV center-of-mass energy. Right: Corresponding total cross
   sections as a function of the neutralino mass. Shown are results at LO (green dotted),
   NLO (blue dashed) and NLO+NLL (red full) together with the corresponding $K$-factors
   and their uncertainties (lower panels). \label{fig:5}}
\end{figure}

Finally, we look at gauginos and their invariant mass distribution
in Fig.\ \ref{fig:5} (left). 
Here, the NLO corrections increase the LO prediction by as much as 30-80\%,
and the NLL contributions add another 6-10\%. The reason is that these
winos are taken to be considerably heavier than the higgsinos, so that
their production occurs closer to threshold, inducing larger logarithmic
corrections. For this reason also the PDF uncertainties are larger in this case.
The total gaugino cross sections are shown in Fig.\ \ref{fig:5} (right)
as a function of the gaugino mass.
Here we observe only a small effect from the resummation in the PDFs.

\section{Conclusion}

To summarize, the public code RESUMMINO now allows to produce predictions at
NLO+NLL for electroweak SUSY particle production using also resummed PDFs.
Although one would in principle expect this to lead to smaller PDF uncertainties,
the fact that these PDFs had to be obtained with a reduced data set leads instead
to larger PDF uncertainties. However, the advantages of resummation corrections
in the PDFs and large data sets entering global PDF analyses can be combined by
using a method based on cross section ratios, i.e.\ $K$-factors.

Large cross sections were predicted in particular for left-handed sleptons
and light higgsinos, while they are naturally smaller for right-handed sleptons
and heavier winos. For the latter, we found instead large resummation corrections
close to threshold. In general, resummation corrections in the matrix elements
were found to be partially compensated by those in the PDFs. Looking ahead,
larger sensitivity and increased cross sections will be provided at the
high-luminosity runs and high-energy upgrade of the LHC \cite{CidVidal:2018eel}.

\end{document}